\documentclass[11pt]{article}

\usepackage[utf8]{inputenc}
\usepackage[T1]{fontenc}

\usepackage{amsmath,amssymb,amsfonts}
\usepackage{mathtools}
\usepackage{physics}
\usepackage{geometry}
\usepackage{commath}
\usepackage{setspace}
\usepackage{mathrsfs}
\usepackage[cal=boondox]{mathalfa}
\usepackage{tikz}
\usepackage{pgfplots}
\pgfplotsset{compat=1.18}
\usetikzlibrary{decorations.markings,arrows.meta}
\usepackage[dvipsnames]{xcolor}

\usepackage[sort&compress,numbers]{natbib}
\usepackage[colorlinks=true,
            linkcolor=blue,
            citecolor=blue,
            urlcolor=blue]{hyperref}

     \usepackage{cancel}
       
\title{\textbf{On the Role of the Canonical Transformation \\
in the Single-Channel Kondo Model}}

\author{
    Zehra Özcan \\
    \textit{Department of Physics, Boğaziçi University} \\
   \textit{34342 Bebek, Istanbul, Türkiye } \\
    \texttt{zehraozcan257@gmail.com}
}

\date{}  

\begin{document}
\maketitle

\begin{abstract}
This pedagogical work presents the significant role that canonical transformation plays in the interpretation of the Abelian bosonized single-channel SU(2) Kondo model, emphasizing its effect on the scaling dimension $\Delta$. The transformation shifts the longitudinal exchange coupling and modifies the scaling dimension of the spin-flip vertex $\tau_{\pm} e^{\pm i\beta \phi}$. Rather than fixing $\Delta$ to the fermionic value $\tfrac{1}{2}$, we keep $\alpha$ explicit, which allows us to identify how different choices lead to marginal or relevant regimes through $(1-\Delta(\alpha))J_\perp$. This approach offers a direct way to trace the scaling behavior from the bosonized Hamiltonian and shows how the RG flow connects to the definition of the Kondo temperature, where the resistance diverges, without switching to other methods.
\end{abstract}

\section{Introduction}

The Kondo model is one of the most important examples in quantum impurity physics. It describes how a single localized spin interacts with conduction electrons, and behaves at low temperatures. Even though it seems simple, the model exhibits several deep ideas in condensed matter physics, such as strong-coupling behavior, and has therefore become a standard test case for many theoretical methods. It was first introduced in the 1960s to explain the unexpected rise in the electrical resistivity of metals that contain a small number of magnetic impurities. In his original paper, Kondo showed that spin-exchange processes between an impurity and the surrounding electrons lead to a logarithmic correction to the scattering amplitude \cite{10.1143/PTP.32.37}. This correction grows as the temperature decreases, meaning that the usual weak-coupling expansion breaks down at low energies. The effect became known as the Kondo effect and motivated new ways to study strongly correlated systems. In 1965, Abrikosov clarified the relation between the resistance and the cutoff scale \cite{PhysicsPhysiqueFizika.2.5}.

In the following decades, several major theoretical methods were developed to understand this behavior more precisely. Anderson’s poor-man’s scaling argument \cite{PWAnderson_1970} gave the first renormalization-group (RG) picture of how the coupling grows and flows to strong coupling. Haldane showed \cite{PhysRevLett.40.911.2}how this scaling picture connects the Anderson and Kondo models and how the Kondo temperature appears as the natural low-energy scale of the problem. Later, the model was solved exactly using the Bethe Ansatz \cite{PhysRevLett.45.379,1980JETPL..31..364V}. Wilson’s numerical renormalization group (NRG) \cite{RevModPhys.47.773} provided a fully nonperturbative way to compute thermodynamic quantities. In parallel, bosonization and conformal field theory methods \cite{PhysRevB.46.10812,PhysRevB.48.7297} uncovered the universal boundary structure of the low-energy fixed point.

In this work, we revisit the single-channel SU(2) Kondo model using Abelian bosonization. The aim is not to introduce new results, but to clarify a step that is often assumed in the literature yet rarely derived in detail: the effect of the canonical transformation on the scaling dimension of the transverse spin-flip term.

In many standard treatments of the Kondo model, the canonical transformation is introduced mainly as a technical route to reach the Toulouse or Emery--Kivelson limit. In these approaches, the scaling dimension $\Delta$ is typically fixed to the fermionic value, and the dependence of the RG relevance of $J_{\perp}$ on the canonical parameter $\alpha$ is stated but not derived explicitly.

Here, we keep $\alpha$ general throughout the bosonization and canonical transformation, making the resulting shift in scaling dimension,
\[
    \Delta(\alpha) = \frac{(\sqrt{2} - \alpha)^2}{2},
\]
transparent and directly connected to the RG flow through $(1 - \Delta(\alpha))\, J_{\perp}$. 
Finally, we examine the Kondo temperature,
\[
k_B T_K \sim D\, e^{-1/(2 \rho J)},
\]
and discuss how its dependence on the transverse coupling can be understood from the scaling dimension $\Delta(\alpha)$ and the bosonized Hamiltonian.  Section~2 introduces the bosonization, Section~3 applies the canonical transformation, and Section~4 interprets the resulting RG structure.

\section{Bosonization of the Single-Channel Kondo Model}

The Hamiltonian for the Kondo model can be separated into two parts,
\begin{align}
    H = H_0 + H_K,
\end{align}
where $H_0$ is the free part and $H_K$ is the interaction part. Writing them in explicit form,
\begin{align}
    H_0 = v_F\sum_{\alpha=1}^{n} \int_{-\infty}^{\infty} 
    \psi_\alpha^\dagger(x)\,(-i\partial_x)\,\psi_\alpha(x)\, dx,
    \qquad
    H_K = \sum_{\nu=1}^{n^2-1} J_\nu\, S^\nu \tau^\nu,
    \label{eq:h0initial}
\end{align}
where $\psi_\alpha$ are fermion fields and $v_F$ is the Fermi velocity. We take $\hbar=1$.  
For the single-channel Kondo model, $n=2$, which gives
\begin{align}
    H_K = \sum_{\nu=1}^{3} J_\nu\, S^\nu \tau^\nu,
\end{align}
where $\tau^\nu$ are $su(2)$ impurity-spin operators, $S^\nu$ are spin densities, and $J_\nu$ are the exchange couplings.

The spin-density operator is defined as
\begin{align}
    S = \sum_{\alpha,\beta=1}^{2} 
    \psi_\alpha^\dagger(0)\,\vec{\sigma}_{\alpha\beta}\,\psi_\beta(0).
\end{align}

\subsection*{Anisotropic Case}
The model becomes anisotropic if the exchange couplings satisfy
\begin{align}
    J_x = J_y \ne J_z.
\end{align}
Historically, $J_x $ \& $ J_y$ are denoted by $J_{\perp}$ and $J_z$ by $J_{\parallel}$, though we keep the notation $J_z$ here.

The anisotropic interaction is written as
\begin{align}
    H_K = H_K^{z} + H_K^{\perp},
\end{align}
with
\begin{align}
H_K^{\perp} &= \frac{J_{\perp}}{2}
\sum_{\alpha,\beta=1}^{2} \sum_{\nu=1}^{2}
\tau_\perp^\nu\, \psi_\alpha^\dagger(0)\,
(\sigma_\nu)_{\alpha\beta}\,\psi_\beta(0)
= \frac{J_{\perp}}{2}
(\tau_{-}S^{+}+\tau_{+}S^{-}),\\
H_K^{z} &= J_{z} \sum_{\alpha,\beta=1}^{2} 
\tau_z\, \psi_\alpha^\dagger(0)\,(\sigma_3)_{\alpha\beta}\,\psi_\beta(0)
= J_{z}\,  (\tau_{z}S^{z}).
\end{align}

\subsection*{Representation of the Fields}
In the Weyl representation, the fermion fields are expressed as
\begin{align}
    \psi = \begin{pmatrix}
        \psi_L \\
        \psi_R
    \end{pmatrix},
\end{align}
where $\psi_L$ and $\psi_R$ denote left-moving and right-moving components respectively.  
Since they can be treated similarly, and in one dimension the left- and right-moving sectors decouple, we focus only on the right-moving case.

\subsection*{Bosonization}
To express the fermion fields in terms of bosonic fields, we apply standard bosonization.  
The spinful fermion fields are bosonized as
\begin{align}
    \psi_{R,s}(x)=\psi_s(x)=\frac{1}{\sqrt{2 \pi a_0}} 
    e^{i\sqrt{4\pi}\,\phi_s(x)},
\end{align}
where $\phi_s(x)$ denotes the spin bosonic field and $a_0$ is a short-distance cutoff. We drop the index $R$.

The bosonic fields separate into charge and spin components,
\begin{align}
    \phi_s = \frac{\phi_{\uparrow}-\phi_{\downarrow}}{\sqrt{2}}, 
    \qquad
    \phi_c = \frac{\phi_{\uparrow}+\phi_{\downarrow}}{\sqrt{2}}.
\end{align}
The impurity couples only to the spin sector, so we focus solely on the spin part of the bosonized fermion fields.

The spin densities are
\begin{align}
S^+ = \frac{1}{2\pi a_0} e^{-i\sqrt{8\pi}\,\phi(x)}, 
\quad
S^- = \frac{1}{2\pi a_0} e^{+i\sqrt{8\pi}\,\phi(x)}, 
\quad 
S^z = \frac{1}{\sqrt{2\pi}}\,\partial_x \phi(x).
\end{align}
Inserting these into the Hamiltonian yields
\begin{align}
    H &= H_0 + H_K^{\perp} + H_K^{z}, \\
      &= H_0 + \frac{J_{\perp}}{2} (\tau_{-}S^{+}+\tau_{+}S^{-})
      + J_{z}\, \frac{\tau_{z}}{\sqrt{2\pi}}\,\partial_x \phi(0), \\
      &= H_0 + \frac{J_{\perp}}{4\pi a_0} 
      (\tau_{+}e^{+i\sqrt{8\pi}\,\phi(0)}+\text{H.c.})
      + J_{z}\, \frac{\tau_{z}}{\sqrt{2\pi}}\,\partial_x \phi(0).
      \label{eq:ham}
\end{align}

\section{Canonical Transformation}

In this section, we now apply the canonical transformation to the bosonized Hamiltonian.
We define the unitary operator
\begin{align}
    U = e^{i\sqrt{4\pi}\alpha \tau_{z}\phi(0)}.
    \label{eq:op-u}
\end{align}
This operator commutes with $\tau_{z}$ but not with $\tau_{\perp} = \tau_{+} + \tau_{-}$.  
Applying the unitary transformation to $\tau_{+}$ gives
\begin{align}
   U^\dagger \tau_{+} U
   = e^{-i\sqrt{4\pi}\alpha \tau_{z}\phi(0)}\, \tau_{+}\,
   e^{i\sqrt{4\pi}\alpha\tau_{z}\phi(0)}.
   \label{eq:unitary}
\end{align}

The Baker–Campbell–Hausdorff (BCH) formula is given by
\begin{align}
    e^A B\,e^{-A} = B + [A,B] + \frac{1}{2}[A,[A,B]] + \frac{1}{3!}[A,[A,[A,B]]] + \cdots
\end{align}
Substituting \eqref{eq:op-u} into \eqref{eq:unitary} and applying the BCH expansion, we find
\begin{align}
     U^\dagger \tau_{+}U
     &= \tau_{+} + [(-i\sqrt{4\pi}\alpha \tau_{z}\phi(0)), \tau_{+}]
     + \frac{1}{2} \Big[(-i\sqrt{4\pi}\alpha \tau_{z}\phi(0)), [(-i\sqrt{4\pi}\alpha \tau_{z}\phi(0)), \tau_{+}] \Big].
     \label{eq:comm}
\end{align}
The spin operators satisfy the commutation relations
\begin{align}
    [\tau_{\pm},\tau_{z}] = \mp\tau_{\pm}, \qquad [\tau_{+},\tau_{-}] = 2\tau_{z}.
\end{align}
Writing \eqref{eq:comm} explicitly,
\begin{equation}
\begin{aligned}
     U^\dagger \tau_{+}U
     &= \tau_{+} + i\sqrt{4\pi}\,\alpha\,\phi(0)[\tau_{+},\tau_{z}]
     + \frac{1}{2} \, 4\pi\alpha^2\,\phi(0)^2 \big[ \tau_{z},[\tau_{+},\tau_{z}] \big]
     + \frac{1}{3!}\#\,[\tau_{z},[\tau_{z},[\tau_{+},\tau_{z}]]] + \cdots \\
     &= \tau_{+} - i\sqrt{4\pi}\,\alpha\,\phi(0)\tau_{+}
     - \frac{1}{2} \, 4\pi\alpha^2\,\phi(0)^2\tau_{+}
     + \frac{1}{3!}\#\,\tau_{+} + \cdots \\
     &= \tau_{+} \, e^{-i\sqrt{4\pi}\alpha\,\phi(0)}.
\end{aligned}
\end{equation}

Substituting this into \eqref{eq:ham}, the perpendicular component of the Hamiltonian becomes
\begin{align}
    H^{\perp} = \frac{J_{\perp}}{4 \pi a_0}
    \Big( \tau_{+} e^{\,i\sqrt{4\pi}(\sqrt{2}-\alpha)\,\phi(0)} + \text{H.c.} \Big).
\end{align}

Next, we apply the canonical transformation to $\partial_x \phi(x)$ using
\begin{align}
    [\phi(x),\partial_x \phi(y)] = -\frac{i}{2}\delta(x-y).
\end{align}
Applying this relation, we obtain
\begin{align}
     U^\dagger \partial_x \phi(x) U
     &= \partial_x \phi(x) - \sqrt{\pi}\,\alpha\, \tau_{z}\,\delta(-x)
     + \frac{1}{2}[\phi(0),\delta(-x)] + \cdots \\
     &= \partial_x \phi(x) - \sqrt{\pi}\,\alpha\, \tau_{z}\,\delta(-x) + 0.
\end{align}
All higher-order terms vanish since $[\phi(0),\delta(x)]=0$.  
Using the fact that the Dirac delta is even,
\begin{align}
    U^\dagger \partial_x \phi(x) U 
    = \partial_x \phi(x) - \sqrt{\pi}\,\alpha\, \tau_{z}\,\delta(x).
\end{align}

Focusing on the local point $x=0$,
\begin{align}
    H^{z} = J_{z} \frac{\tau_{z}}{\sqrt{2\pi}}
    \Big( \partial_x \phi(0)
    - \sqrt{\pi}\,\alpha\,\tau_{z} \int dx\,\delta(x) \Big),
\end{align}
where $\tau_z$ is a spin-$\frac{1}{2}$ operator, so $\tau_z^2 = \frac{1}{4}$.  
Integrating the Dirac delta yields a boundary correction, giving
\begin{align}
    H^{z} = \frac{J_{z}}{\sqrt{2\pi}}
    \Big( \tau_{z} \partial_x \phi(0) - \frac{\sqrt{\pi}\,\alpha}{4} \Big).
    \label{eq:zcom}
\end{align}

Finally, we apply the transformation to $H_0$, as in the previous steps
\begin{align}
      U^\dagger H_0 U = v_F \int dx \big( U^\dagger \partial_x \phi(x) U \big)^2.
      \label{eq:h0canonical}
\end{align}
Inserting the expression obtained above,
\begin{align}
      U^\dagger H_0 U
      &= v_F \int dx \Big( \partial_x \phi(x)
      - \sqrt{\pi}\,\alpha\, \tau_{z}\,\delta(x) \Big)^2 \\
      &= H_0 + v_F \int dx 
      \Big( \pi\alpha^2 \tau_z^2 (\delta(x))^2
      - 2\tau_z \partial_x \phi(x)\sqrt{\pi}\,\alpha\,\delta(x) \Big).
\end{align}
Since the term with $ (\delta(x))^2$  commutes with all dynamical operators, it does not contribute to the equations of motion and is therefore dropped.
\begin{align}
  U^\dagger H_0 U  &= H_0 
      - 2\sqrt{\pi}\,v_F\,\alpha\,\tau_z \partial_x \phi(0).
      \label{eq:h0}
\end{align}
Combining \eqref{eq:h0} and \eqref{eq:zcom}, we obtain
\begin{align}
    H'_0 &= H_0  - \frac{J_{z}\alpha}{4\sqrt{2}}, \\
    H^{z'} &= \frac{J_{z}\tau_{z}\partial_x \phi(0)}{\sqrt{2\pi}}
    - 2\sqrt{\pi}v_F\,\alpha\,\tau_z \partial_x \phi(0).
\end{align}

\section{Renormalization Group Flow and Scaling Analysis}

After the canonical transformation, collecting all terms, the effective Hamiltonian can be written as
\begin{equation}
H' = H_0'[\phi] + 
\frac{J_\perp}{2\pi a_0}
\bigl[\tau_+ e^{i\sqrt{4\pi}(\sqrt{2}-\alpha)\phi(0)} + \text{H.c.}\bigr]
+ \frac{\lambda}{\sqrt{\pi}}\,\tau_z\,\partial_x\phi(0),
\end{equation}
where
\begin{align}
    \lambda = \frac{J_z}{\sqrt{2}} - 2\pi v_F \alpha.
    \label{eq:lambda}
\end{align}
This shows that the canonical transformation effectively shifts the longitudinal exchange by an amount proportional to $\alpha$, while modifying the exponent of the spin–flip term.

The scaling dimension is defined by
\begin{align}
    \Delta = \frac{\beta^2}{8\pi},
\end{align}
where $\beta$ comes from $e^{i\beta \phi(0)}$. From $H'$ above, we identify
\begin{align}
    \beta = \sqrt{4\pi}(\sqrt{2}-\alpha),
\end{align}
giving
\begin{equation}
\Delta = \frac{(\sqrt{2}-\alpha)^2}{2}.
\label{eq:scaling}
\end{equation}

Near the ultraviolet fixed point, the RG flow of the couplings is 
\begin{align}
\frac{dJ_\perp}{d\ell} &= (1-\Delta)J_\perp,  \label{eq:jperp}\\
\frac{dJ_z}{d\ell} &= c J_\perp^2,
\end{align}
where the first equation follows at tree level from the scaling dimension of the boundary vertex operator\cite{Giamarchi2003ooa}, while the second one is from poor man’s scaling \cite{PWAnderson_1970, PhysRevLett.40.911.2}. Here $\ell = \ln(D_0/D)$ with $c > 0$ a constant, and $D$ the conduction electron bandwidth.

%

\subsection{Varying the $\alpha$}

Assuming positive coupling constants ($J_{\perp,z}>0$), the term $(1 - \Delta)$ determines whether the transverse term is relevant. From $\Delta(\alpha) = \frac{(\sqrt{2}-\alpha)^2}{2}$, we see that 
keeping $\alpha$ explicit lets us examine different regimes.

Distinct values of $\alpha$ correspond to:
\begin{itemize}
    \item $\alpha = \sqrt{2} - 1$ ($\Delta = 1/2$): This is the \textit{fermionic} scaling as it matches that of a single fermion operator. This allows us to perform refermionization. Together with $\lambda = 0$, it gives the Toulouse (Emery–Kivelson) limit where the model is exactly solvable. The perturbation is relevant.
    \item $\alpha = 0$ ($\Delta = 1$): This corresponds to the \textit{bosonic} scaling. The perturbation is marginal, since \eqref{eq:jperp} vanishes at leading order and one must consider higher-order terms. The flow of $J_\perp$ is logarithmically slow.
    \item $\alpha = \sqrt{2}$ ($\Delta = 0$): The exponential factors in the Hamiltonian simplify to unity. In this case the transverse perturbation is maximally relevant. The coupling flows very rapidly to strong coupling. 
\end{itemize}

For $0 \leq \alpha < \sqrt{2}$, we have $\Delta < 1$ and the transverse term $J_\perp$ is relevant and drives the system toward the strong–coupling screened fixed point. At $\alpha = 0$ the perturbation is marginal at tree level but becomes marginally relevant for antiferromagnetic $J_z>0$, causing the usual Kondo flow. Increasing $\alpha$ lowers the scaling dimension and strengthens this relevance.

Since $\Delta(\alpha)=(\sqrt{2}-\alpha)^2/2$ is defined through a square, the condition $\Delta>1$ would require $\alpha<0$ or $\alpha>2\sqrt{2}$. In such a case the perturbation is irrelevant and the coupling constant $J_\perp$ flows to zero.

As $\alpha$ varies, the scaling behavior changes. In other words, the canonical transformation is not just a technical step but a useful way to examine how the scaling dimension changes across different limits through a single algebraic parameter.

\subsection{Ferromagnetic and Antiferromagnetic Cases}
We can also consider two cases depending on the sign of longitudinal exchange coupling $J_z$.
\begin{itemize}
    \item \textit{Ferromagnetic interaction $(J_z < 0)$:} The coupling constants approach zero as the conduction electron bandwidth cutoff ratio, $\frac{D_0}{D}$, increases.
    \item \textit{Antiferromagnetic interaction $(J_z > 0)$:} The system goes towards strong coupling as the conduction electron bandwidth cutoff ratio $\frac{D_0}{D}$ increases, ($D\xrightarrow{}0$), meanwhile the coupling constants go to infinity.
\end{itemize}

\section{Interpretation of Temperature Through RG}

Kondo \cite{10.1143/PTP.32.37} derived the temperature dependence of the resistance using the Born approximation:
\begin{align}
    R = R_0 \left[1 - 4J \rho \ln\left(\frac{k_B T}{D}\right) + \dots \right],
\end{align}
where $\rho$ is the density of states at the Fermi level.  
In the following year, Abrikosov resummed the logarithmic terms and obtained \cite{PhysicsPhysiqueFizika.2.5}
\begin{align}
    R = \frac{R_0}{\left(1 + 2J \rho \ln\left(\frac{k_B T}{D}\right)\right)^2}.
\end{align}
From this expression, the Kondo temperature is defined as the temperature at which the resistance diverges:
\begin{align}
    k_B T_K \sim D\, e^{-1/(2 \rho J)}.
\end{align}

\vspace{0.5em}
To connect this with the RG flow, we integrate the transverse coupling from
\[
\frac{dJ_\perp}{d\ell} = (1 - \Delta) J_\perp,
\]
which yields
\begin{align}
  J_\perp(D) = A \left(\frac{D_0}{D}\right)^{1-\Delta},
\end{align}
where we denote $A \equiv J_\perp(D_0)$.  
Substituting into the expression for $T_K$ gives
\begin{align}
    k_B T_K \sim D\, \exp\!\left[
    -\frac{1}{2 \rho A} \left(\frac{D}{D_0}\right)^{1-\Delta}
    \right].
\end{align}

This expression shows how the scaling dimension $\Delta(\alpha)$ directly governs the
temperature dependence of the effective exchange coupling, providing a bridge
between the algebraic RG structure of the bosonized Hamiltonian and the
thermodynamic behavior of the Kondo effect.

For $\Delta < 1$, the exponent becomes negative, indicating that the
transverse coupling grows as $T$ decreases. This corresponds to the strong–coupling regime where the impurity becomes screened.

While $\Delta=0$ corresponds to $\alpha \to \sqrt{2}$, where the vertex operator in the bosonized theory becomes non-normalizable. However, the RG equations and resulting temperature dependence remain analytic.  In this limit, the coupling is
maximally relevant, leading to an immediate flow toward the strong-coupling
fixed point.  The Kondo temperature thus remains finite and well-defined.

\section{Discussion}
In most treatments of the Kondo model, the canonical transformation is introduced mainly to reach the Toulouse limit. However, its effect on the scaling dimension is typically assumed rather than derived explicitly. This note attempts to present a step-by-step derivation of how varying $\alpha$ modifies the vertex exponent and hence the RG behavior.

Keeping the canonical transformation parameter $\alpha$ explicit throughout the calculation makes it possible to see how it shifts the longitudinal exchange term and changes the scaling dimension of the transverse spin-flip term. Instead of fixing $\alpha$ at one special value, this allows us to explore how different choices correspond to different physical regimes.

The scaling dimension $\Delta(\alpha)$ determines how the transverse coupling $J_\perp$ behaves under renormalization. When $\Delta < 1$, the coupling grows at low energies, meaning that spin-flip scattering becomes stronger and drives the system toward strong coupling. At the Toulouse point, the longitudinal term is removed by the transformation, and the Hamiltonian becomes exactly solvable. Furthermore, the RG flow of $J_\perp$ can be connected to the temperature dependence of the resistance through the definition of the Kondo temperature $T_K$.  This shows that the certain physical features of the model can be approached from the bosonized Hamiltonian.


\section*{Acknowledgments}
I would like to thank Prof.\ Teoman Turgut for his guidance and support during this work.
This note grew out of my attempts to understand the bosonized form of the Kondo model during my undergraduate studies. 
It is not intended to present new results, but rather to work through a derivation that I initially found difficult to access in the literature. 
Any remaining misunderstandings are my own.

\appendix
\section{Free Hamiltonian}

Following Ref.~\cite{sénéchal1999introductionbosonization}, we obtain
Eq.~\eqref{eq:h0canonical} from Eq.~\eqref{eq:h0initial}.
\begin{align}
    H_0= v_F \int dx \, \psi^{\dagger}(x)(-i\partial_x )\psi(x)
\end{align}
Since we are dealing with vertex operators, normal ordering is required.
This can be implemented either by mode expansion or by point splitting.
Here, we use the latter method since it is shorter.
\begin{align}
   (-i) \psi^{\dagger}(x)\partial_x \psi(x)
   =(-i) \lim_{\epsilon \rightarrow0}
   [\psi^{\dagger}(x+ \epsilon)\partial_x \psi(x)
   - \langle\psi^{\dagger}(x+ \epsilon)\partial_x \psi(x) \rangle]
\end{align}
Applying bosonization, we obtain
\begin{align}
    =\frac{-i}{2 \pi a_0}\lim_{\epsilon \rightarrow0}
    [ e^{-i\sqrt{4\pi}\,\phi(x+\epsilon)}\partial_xe^{i\sqrt{4\pi}\,\phi(x)}
    -\langle e^{-i\sqrt{4\pi}\,\phi(x+\epsilon)}\partial_xe^{i\sqrt{4\pi}\,\phi(x)}\rangle ].
\end{align}
Focusing on the correlation function, we write
\begin{align}
    \langle e^{-i\sqrt{4\pi}\,\phi(x+\epsilon)}\partial_xe^{i\sqrt{4\pi}\,\phi(x)}\rangle
    =(i\sqrt{4\pi})\langle
    e^{-i\sqrt{4\pi}\,\phi(x+\epsilon)}
    e^{i\sqrt{4\pi}\,\phi(x)} \partial_x\phi(x)\rangle.
\end{align}

This can be calculated using the mode expansion~\cite{sénéchal1999introductionbosonization}:
\begin{align}
    \phi(x)
&= Q + \frac{P}{2L }(vt - x)
   + \sum_{n>0} \frac{1}{\sqrt{4\pi n}}
   \left( b_n e^{-ik_n (vt - x)} + b_n^\dagger e^{ik_n (vt - x)} \right).
\end{align}
Then,
\begin{equation}
\begin{aligned}
\big\langle \phi(x)\phi(x') \big\rangle
&= \Big\langle
\Big[
Q + \frac{P}{2L }(vt - x)
+ \sum_{n>0} \frac{1}{\sqrt{4\pi n}}
  \left( b_n e^{-ik_n (vt - x)} + b_n^\dagger e^{ik_n (vt - x)} \right)
\Big]
\\
&\qquad\qquad \times
\Big[
Q + \frac{P}{2L }(vt - x')
+ \sum_{m>0} \frac{1}{\sqrt{4\pi m}}
  \left( b_m e^{-ik_m (vt - x')} + b_m^\dagger e^{ik_m (vt - x')} \right)
\Big]
\Big\rangle .
\end{aligned}
\end{equation}
This can be written as
\begin{align}
\big\langle \phi(x)\phi(x') \big\rangle
= \langle \varphi(x)\varphi(x') \rangle
\;+\; \text{(zero-mode pieces)} .
\end{align}
where
\begin{align}
\text{zero-mode pieces} &=
\langle Q^2 \rangle
+ \frac{(vt - x)(vt - x')}{4 L^2} \langle P^2 \rangle
+ \frac{vt - x}{2L } \langle QP \rangle
+ \frac{vt - x'}{2L } \langle PQ \rangle .
\end{align}
By normal ordering, the zero-mode pieces can be dropped. Thus,
\begin{align}
\big\langle \phi(x)\phi(x') \big\rangle
= \langle \varphi(x)\varphi(x') \rangle.
\end{align}
Focusing on the terms involving the creation and annihilation operators, we find
\begin{equation}
\begin{aligned}
\langle \varphi(x)\varphi(x') \rangle
&= \Big\langle 0 \Big|
\sum_{n>0}\frac{1}{\sqrt{4\pi n}} 
 \bigl[b_n e^{-ik_n(vt-x)}+b^\dagger_n e^{ik_n(vt-x)}\bigr]\\
&\qquad \times \sum_{m>0}\frac{1}{\sqrt{4\pi m}}
\bigl[b_me^{-ik_m(vt-x')}+b^\dagger_m e^{ik_m(vt-x')}\bigr] \vphantom{\sum_{n>0}} \Big| 0 \Big\rangle.
\end{aligned}
\end{equation}

Using the commutation relation
$ [b_n,b^\dagger_m]=\delta_{nm}$ and the quantized wavevector equality
$k=\frac{2\pi n}{L }$, we obtain
\begin{align}
    &= \sum_{n>0}\sum_{m>0}\,
\frac{1}{4\pi\sqrt{mn}}\,e^{-ik_n(vt-x)}e^{ik_m(vt-x')}
\bigl\langle 0\big| b_n b^\dagger_m\big|0\bigr\rangle, \\
&= \sum_{n>0}\,\frac{1}{4\pi n}\,e^{-ik_n(x-x')},
\qquad (x>x') \\
&= \sum_{n>0}\,\frac{1}{4\pi n}\,e^{-\frac{2\pi i n}{L}(x-x')}.
\end{align}
The series is summed using
\begin{align}
    \sum_{n>0} \frac{e^{- \alpha n}}{n}=- \ln{(1-e^{- \alpha})},
\end{align}
which gives
\begin{align}
    \langle \phi(x)\phi(x') \rangle
=\frac{-1}{4\pi }\ln{(1-e^{-\frac{2\pi i }{L}(x-x')})}
\end{align}

Since $|x-x'| \ll L$, we expand the exponential to leading order,
\begin{align}
1 - e^{-\frac{2\pi i}{L}(x-x')} \simeq \frac{2\pi i}{L}(x-x').
\end{align}
This gives
\begin{align}
\langle \phi(x)\phi(x') \rangle
= -\frac{1}{4\pi}\ln\!\left(\frac{2\pi i (x-x')}{L}\right).
\end{align}
We reintroduce a short-distance cutoff $a_0$,
\begin{align}
\langle \phi(x)\phi(x') \rangle
= -\frac{1}{4\pi}\ln\!\left(\frac{2\pi i (x-x')}{a_0}\frac{a_0}{L}\right).
\end{align}
Taking the thermodynamic limit $L \to \infty$,
\begin{align}
\ln\!\left(\frac{2\pi i (x-x')}{a_0}\frac{a_0}{L}\right)
= \ln\!\left(\frac{2\pi i (x-x')}{a_0}\right)
+ \ln\!\left(\frac{a_0}{L}\right).
\end{align}
The second term is independent of $x-x'$ and can therefore be absorbed into an additive constant. We finally obtain
\begin{align}
\langle \phi(x)\phi(x') \rangle
= -\frac{1}{4\pi}\ln\!\left(\frac{2\pi i (x-x')}{a_0}\right).
\end{align}

For a Gaussian field, the exponential of a linear functional satisfies
\begin{align}
    \Big\langle
      e^{i\alpha\phi(x)} e^{-i\alpha\phi(x')}
    \Big\rangle
    = \exp\!\big(\alpha^2 \langle\phi(x)\phi(x')\rangle\big).
\end{align}
When \(x-x'=\epsilon\), the logarithm contains \(\ln(i\epsilon)\), which can be taken with two opposite signs depending on the branch. We choose the branch that leads to a positive free Hamiltonian, which gives
\begin{align}
    C(\epsilon)
    \equiv
    \Big\langle
      e^{-i\sqrt{4\pi}\,\phi(x+\epsilon)}
      e^{i\sqrt{4\pi}\,\phi(x)}
    \Big\rangle
    =- \frac{a_0 }{2\pi i\epsilon}.
\end{align}
This is consistent with the known right-moving fermion correlation function,
\begin{align}
    \langle \psi^{\dagger}(x)\psi(x')\rangle=\frac{1}{2\pi}\frac{1}{x-x'}
\end{align}

Thus, the expectation value of the point–split operator is
\begin{align}
    \Big\langle
      \psi^{\dagger}(x+ \epsilon)\partial_x \psi(x)
    \Big\rangle
    = i\sqrt{4\pi}\,
      C(\epsilon)\,\partial_x\phi(x).
\end{align}

Moreover, using the BCH identity,
\begin{align}
    e^{i \alpha \phi(x)}e^{i \beta \phi(x')}
    =e^{i \alpha \phi(x)+i \beta \phi(x')}
    e^{-\alpha \beta \langle\phi(x)\phi(x')\rangle}
\end{align}
we obtain
\begin{align}
    \psi^{\dagger}(x+ \epsilon)\partial_x \psi(x)
    &= i\sqrt{4 \pi}
    (e^{-i\sqrt{4\pi}\,\phi(x+\epsilon)}
    e^{i\sqrt{4\pi}\,\phi(x)})
    \partial_x \phi(x) ,\\
    &= i\sqrt{4\pi} \frac{(-a_0) }{2\pi i\epsilon}
    e^{-i\sqrt{4\pi}(\phi(x+\epsilon)-\phi(x))}
    \partial_x \phi(x), \\
    &=-\frac{a_0}{\sqrt{\pi}\epsilon}
    e^{-i\sqrt{4\pi}(\epsilon\partial_x \phi(x))}
    \partial_x \phi(x).
\end{align}
Substituting into the point–split expression,
\begin{align}
    \lim_{\epsilon \rightarrow0}
    [\psi^{\dagger}(x+ \epsilon)\partial_x \psi(x)
    - \langle\psi^{\dagger}(x+ \epsilon)\partial_x \psi(x) \rangle]
    =\frac{-i}{2 \pi a_0}
    \lim_{\epsilon \rightarrow0}
    \left[\frac{-a_0}{\sqrt{\pi}\epsilon}
    e^{-i\sqrt{4\pi}(\epsilon\partial_x \phi(x))}
    \partial_x \phi(x)
    + i\sqrt{4\pi}\frac{a_0 }{2\pi i \epsilon}\partial_x\phi(x)\right].
\end{align}
Taylor expanding the exponential,
\begin{align}
    =&\frac{-i}{2 \pi a_0}\lim_{\epsilon \rightarrow0}
    \left[ \frac{-a_0}{\sqrt{\pi}\epsilon}
    (\cancel{1}-i \epsilon \sqrt{4\pi}\partial_x \phi(x))
    \partial_x\phi(x)
    +\cancel{\frac{a_0}{\sqrt{\pi}\epsilon}\partial_x\phi(x)}\right], \\
    =&\frac{-i}{2 \pi a_0}\lim_{\epsilon \rightarrow0}
    \left[i \, a_0 \sqrt{4} (\partial_x \phi(x))^2\right], \\
    =&\frac{1}{\pi }(\partial_x\phi(x))^2.
\end{align}
After rescaling the field, this yields
\begin{align}
    H_0
    = v_F \int dx \, (\partial_x\phi(x))^2,
\end{align}
which is Eq.~\eqref{eq:h0canonical}.

\nocite{*}
\bibliography{ref}

\end{document}